\documentclass{article}

\usepackage{chicago}
\usepackage{amssymb}
\usepackage{amsmath}
\usepackage{graphicx}

\newcommand{\iec}{\mbox{i.\,e.\,}}

\newcommand{\egc}{\mbox{e.\,g.\,}}

\newcommand{\dr}[1]{\ensuremath{\mathrm{d} #1\,}}
\newcommand{\mc}[1]{\ensuremath{\mathcal{#1}}}

\newcommand{\ddt}{\ensuremath{\frac{\dr{}}{\dr{t}}}}

\newcommand{\tr}{\textsf{Tr}}

\newcommand{\be}{\begin{equation}}
\newcommand{\ee}{\end{equation}}
\newcommand{\e}[1]{\mathrm{e}^{#1}}

\usepackage[version=4]{mhchem}

\begin{document}

\title{What Gibbsian Statistical Mechanics Says: in defense of bare probabilism}
\author{David Wallace\thanks{Department of History and Philosophy of Science / Department of Philosophy, University of Pittsburgh, PA 15260, USA; email \texttt{david.wallace@pitt.edu}}}
\maketitle

\begin{abstract}I expound and defend the ``bare probabilism'' reading of Gibbsian (i.e. mainstream) statistical mechanics, responding to Frigg and Werndl's recent (\emph{BJPS} 72 (2021), 105-129) plea: ``can somebody please say what Gibbsian statistical mechanics says?''
\end{abstract}

\section{Introduction}

Mainstream statistical physics proceeds by assigning probability functions to classical systems, and mixed quantum states to quantum systems, and then calculating synchronic and diachronic properties of those functions. Recent philosophy of physics refers to this mainstream approach as ``Gibbsian statistical mechanics'' (henceforth GSM) and contrasts it, usually unfavorably, to (so-called) ``Boltzmannian statistical mechanics'', in which the role of probability is lessened and in some versions eliminated altogether.

Recent philosophy of physics, however, has seen increasing interest in engaging with rather than simply dismissing GSM (see, \egc, \citeNP{luczak-equilibrium}, Robertson~\citeyearNP{robertsoncoarsegraining,robertsonholygrail}, \citeNP{myrvoldbook}, Wallace~\citeyearNP{wallaceactualstatmech,wallacegibbsnecessity}), and this has led to controversy about its actual content. The issue has been raised in a recent (and characteristically clear) paper by \citeN{friggwerndlcansomeoneplease} (henceforth FW), titled ``Can somebody please say what Gibbsian statistical mechanics says?'' In their introduction, they say\begin{quote}
GSM is widely used and considered by many to be \emph{the} theory of statistical mechanics. Yet a closer look at GSM reveals that it is unclear what the theory says and how it bears on experimental practice. \ldots Hence our plea: can somebody please says what GSM says? (p.105; emphasis theirs)
\end{quote}
This paper is an attempt to answer their plea (building on a previous, less explicit answer in \cite{wallaceactualstatmech}), as well as a critique of the alternative answer that FW themselves develop. I defend the claim that GSM should be interpreted --- and is interpreted by working physicists --- via what FW call \emph{bare probabilism}, the view that there is \emph{nothing} to GSM beyond the probabilistic claims that it makes about systems. FW regard this as unacceptable in large part because it fails to provide an adequate basis for thermodynamics; I will argue that this follows only because FW fail to recognize that in modern physics, thermodynamics is interpreted statistically just as statistical mechanics is.

In section \ref{formalism} I review the formalism of GSM, and in section \ref{FWcrit} I summarize FW's criticisms and their eventual position. In section \ref{statistical-thermodynamics} I defend the view that contemporary thermodynamics should itself be understood statistically, and in sections \ref{fluctuations}--\ref{equilibration} I discuss fluctuations in GSM, which FW regard as a major issue for its interpretation. Section~\ref{conclusion} is the conclusion.

A disclaimer about terminology: the philosophy literature can easily give the impression that (1) the ``Gibbs-vs-Boltzmann'' debate is a genuine schism within modern physics, and (2) what is called ``Gibbsian'' or ``Boltzmannian'' statistical mechanics is an accurate description of the historical views of (respectively) Gibbs and Boltzmann. I reject both claims. As to the first: I agree with FW that ``Gibbsian statistical mechanics'' is simply what modern physicists call statistical mechanics; ``Boltzmannian statistical mechanics'' can be understood varyingly as a historically held position, as the view of a heterodox minority~\cite{lebowitz,goldsteinboltzmann}, or as a special case of mainstream statistical mechanics~\cite{wallacegibbsnecessity}. As to the second: Wayne Myrvold~\citeyear[ch.7]{myrvoldbook} persuasively argues that the historical Gibbs and Boltzmann held significantly more nuanced, and more mutually compatible, views than the contemporary literature might suggest. I have stuck with the standard terminology for ease of communication, but to be clear: by ``GSM'' I mean no more or less than ``statistical mechanics as it has in fact been practiced in mainstream physics over the last seventy years or so''.

\section{The Gibbsiam formalism}\label{formalism}

The \emph{formalism} of GSM (leaving aside its interpretation for now) is as follows: 
\begin{enumerate}
\item A \emph{system} is characterized either classically by a phase space and a collection of functions on that space representing the dynamical variables (including the Hamiltonian $H$, which generates the system's dynamics) or quantum-mechanically by a Hilbert space and a collection of operators on that space representing the dynamical variables (again including the Hamiltonian $H$).
\item The \emph{statistical state} $\rho(t)$ of the system at time $t$ is defined either classically by a probability measure over the phase space, or quantum-mechanically by a quantum state (pure or mixed) on the Hilbert space. (Notice an important quantum/classical disanalogy: in classical mechanics there is a clear distinction between the statistical state and the \emph{microstates} of the system, which are represented by \emph{points} in the phase space; in quantum mechanics the distinction is elided).
\item The statistical state of an undisturbed system evolves over time under Liouville's equation, which classically is
\be
\ddt{\rho}(t) = \{H,\rho\}_{PB}
\ee
with $\{,\}_{PB}$ the Poisson bracket, and quantum-mechanically as
\be
\ddt{\rho}(t) = -i[H,\rho]
\ee
with $[,]$ the commutator.
\item Given a dynamical variable $X$, its \emph{expectation value} with respect to statistical state $\rho$ is given classically\footnote{Technical note: the integral in (\ref{classicalexpectation}) is with respect to the Liouville measure, and I am assuming the probability measure is represented by a (possibly improper) function $\rho(x)$ defining its local ratio to the Liouville measure. I adopt this notation throughout, following FW; readers who prefer a more explicitly measure-theoretic notation should be able to translate straightforwardly.} by
\be\label{classicalexpectation}
\langle X \rangle_\rho = \int\dr{x}\rho(x)X(x)
\ee
and quantum-mechanically by
\be
\langle X \rangle_\rho =\tr (\rho X).
\ee
\end{enumerate}
A system is in \emph{exact statistical equilibrium} if its statistical state is invariant under time, $\mathrm{d}\rho / \mathrm{d}t=0$. One important class of states in exact statistical equilibrium are the \emph{canonical states}, which have form
\be
\rho(\beta)=\frac{1}{Z(\beta)}\e{-\beta H}
\ee
(interpreted either classically or quantum-mechanically; $Z(\beta)$ is in either case a normalization factor), and a system is in \emph{exact canonical equilibrium} if its statistical state is one of the canonical states. Exact canonical equilibrium is thus a special case of exact statistical equilibrium. The notion of canonical state generalizes naturally to include dependence on other conserved quantities such as $N$, the number of particles; for this case, we have
\be
\rho(\beta,\mu)=\frac{1}{Z(\beta,\mu)}\e{-\beta(H+\mu N)}.
\ee
(States like this are sometimes called \emph{grand canonical states}.)

It is common\footnote{In the philosophy literature FW attribute this definition to \citeN{Lith1999}. It is very widespread in the physics literature: notably, it is the central notion used by the various recent programs to explain equilibration in quantum statistical mechanics (cf section \ref{equilibration}).} to introduce a weaker notion of statistical equilibrium. We fix some set of \emph{macrovariables}, which are a small subset of the degrees of freedom intended to represent collective, large-scale degrees of freedom and/or degrees of freedom which are realistically measureable. Common choices include: for a spatially extended system the microvariables averaged over some lengthscale $L$ large compared to intermolecular spacing but small compared to the system scale (or, more practically, the Fourier components of the dynamical variables with wavelengths $>L$), and for a system of $N$ particles, the complete set of $M$-particle dynamical variables for fixed $M\ll N$. (These roughly correspond to the choices of macrovariables generally made in the two main paradigms for non-equilibrium statistical mechanics: the Langevin and BBGKY paradigms (see, \egc, \cite[ch.2]{calzettahubook} or \cite{wallacequantumstatmech}. In the former paradigm the macrovariables are sometimes called \emph{relevant} variables.)
A system is then in \emph{statistical macroequilibrium} at $t$ if the expectation value $\langle M(t)\rangle\equiv \langle M \rangle_{\rho(t)}$
is time-invariant at $t$ for all macrovariables $M$. And it is in \emph{canonical macroequilibrium} if the expectation values of all macrovariables match their values on some canonical state.

As FW note, this formalism is related to thermodynamics by what they call the \emph{averaging principle} (AP): thermodynamic quantities like energy or particle number are identified with the expectation values of the corresponding statistical system, \emph{calculated at canonical equilibrium}. To spell this out a little:\footnote{For a more detailed description, see \cite[section 3]{wallaceirreversibility} (or any good statistical mechanics textbook).} consider a fluid, which in phenomenological thermodynamics is characterized by its energy $U$,  the number of particles $n$, and its volume $V$; the fluid's thermodynamic behavior is described by its equation of state $S=S(U,n,V)$, where $S$ is the system's thermodynamic entropy. In Gibbsian statistical mechanics we identify $S(U,N,V)$ with the Gibbs entropy of the unique canonical state that has expected energy $\langle H(V) \rangle = U$ and expected number of particles $\langle N\rangle = n$, where the system volume $V$ is interpreted as an external parameter for the Hamiltonian $H(V)$ and where $N$ is the classical function or quantum operator representing particle number. In \cite[section 3]{wallaceirreversibility} I call this fuller recipe (which includes AP as a part of it) the \emph{canonical recipe}: using it, in those situations where it is calculationally tractable to do so, reliably recovers the measured thermodynamic equation of state.

\section{Frigg and Werndl on the interpretation of GSM}\label{FWcrit}

FW raise two, related, questions about GSM:
\begin{enumerate}
\item How is the formalism to be interpreted; that is, what if anything are we saying \emph{about a physical system} when we assign a statistical state to it?
\item How is AP to be justified, given some interpretation of GSM?
\end{enumerate}
(I have stated (2) as secondary to (1), as I think this is faithful to FW, but I would myself be inclined to break it loose from any question of interpretation and ask: why does AP (and, more generally, the canonical recipe) in fact give \emph{empirically correct predictions} for thermodynamic systems?)

They begin by considering, and fairly swiftly rejecting, two possibilities: \emph{quietism} (the view that AP is simply a posit in no need of further justification) and the \emph{time average approach}, which equates the Gibbsian probability measure to a long-term time average. In both cases I agree with their assessment and will not discuss these approaches further.\footnote{Interestingly, FW describe the time average approach as ``the `standard' textbook approach''; if so, I think that reflects badly on the writers of textbooks (or at least, reflects those writers' impatience to get through foundations as quickly as possible so as to discuss applications), since severe criticisms of the approach go back at least as far as \cite{tolman}. \citeN{Penrose1979} described this approach as ``out of fashion now'' over forty years ago.}

FW's main focus is \emph{probabilism}: the view that GSM is to be interpreted in the obvious way as making probabilistic statements about systems, so that in classical mechanics, when we say `the statistical state of this system is $\rho$', we mean, the probability density of the system having microstate $x$ is $\rho(x)$. FW do not discuss quantum mechanics explicitly, but the obvious quantum version of probabilism is just that a system's statistical state is its actual quantum state. (What that means, and in particular whether it is itself a probabilistic or categorical statement, of course depends on one's preferred solution to the measurement problem.)

FW consider several versions of probabilism (as well as a digression to discuss a recent proposal by \citeN{mccoystatmech}, which they ultimately reject; I will not discuss that proposal here). The first is \emph{bare probabilism}: the view that \emph{all there is} to Gibbsian statistical mechanics is the probabilistic interpretation. As FW say, ``[o]n such a view GSM really is just a study of the statistical properties of [the statistical state] with nothing else added.''\footnote{They cite my \citeyear{wallaceactualstatmech} when they present bare probabilism, but say (fn. 16) that they have been unable to find an explicit statement of bare probabilism in print. I had intended to give one; obviously I was insufficiently clear!}

To FW, the problem with bare probabilism is that --- as they interpret it --- it abandons any hope of a statistical-mechanical grounding of thermodynamics. (``Any attempt to read more into GSM, in particular any attempt to read a notion of thermodynamic equilibrium into it, is misguided and should be resisted.'') They go on to spell out the concern in more detail:
\begin{quote}
The fact that thermodynamic equilibrium and statistical equilibrium are both equilibria does not mean that they are somehow similar, or that statistical equilibrium can serve as a stand-in for thermodynamic equilibrium when the latter is excised. In fact, statistical and thermodynamic equilibrium are not only conceptually different, they do not even have the same extension. An ensemble in statistical equilibrium not only contains systems at thermodynamic equilibrium; it can also contain systems that are not at thermodynamic equilibrium.
\end{quote}
Here, FW are making substantive assumptions about thermodynamics that are worth spelling out. Specifically, they are committed to the `neo-Boltzmannian' view that thermodynamic properties like energy, entropy and being at equilibrium are properties of \emph{microstates}. For equilibrium in particular, and given again a collection of macrovariables we can define a microstate of a classical system as being at \emph{Boltzmannian equilibrium} if the value of the macrovariables is time-invariant, perhaps up to some small fluctuation term; FW (section 2) define thermodynamic equilibrium as Boltzmannian equilibrium.

Under these assumptions about thermodynamics, FW are surely correct that thermodynamics does not reduce to GSM given Bare Probabilism, that statistical equilibrium is not thermodynamic equilibrium (it is certainly not Boltzmannian equilibrium), and that AP does not follow.

Given these issues, FW regard bare probabilism as unsatisfactory. They consider two alternatives, both of which have as a starting point a consideration of statistical fluctuations. Staying within classical mechanics, they define the \emph{fluctuation} of a system with microstate $x(t)$ at time $t$ (with respect to some macrovariable $f$) as
\be
\Delta(t) = f(x(t))-\langle f \rangle_\rho
\ee
where $\rho$ is the equilibrium distribution (presumably the canonical or microcanonical distribution, though FW are not explicit here). Then they suggest that AP would be justified as a good approximation for any system where the probability of large $\Delta(t)$ is low (a condition that they call \emph{thermodynamic fluctuations}). How we could generalize this to quantum mechanics is unclear, but FW restrict their attention to classical mechanics.

FW then suggest two modifications of probabilism. In \emph{qualified probabilism}, AP is added as a requirement of inter-theoretic reduction: `bare probabilism must ensure that AP holds whenever GSM is used in tandem with [thermodynamics]'. In \emph{fluctuation probabilism}, AP is a restriction on GSM itself: `[f]luctuation probabilism has to restrict dynamical laws and observables that are allowable in GSM to those that produce thermodynamic fluctuations'. 

FW claim that it is an `aesthetic matter' which of qualified or fluctuation probabilism we accept, but it's not clear why, unless we regard GSM as purely a foundation for thermodynamics: according to qualified probabilism we can use GSM outside the regimes where AP holds, so long as we don't try to relate it to thermodynamics; according to fluctuation probabilism we cannot use GSM at all unless AP holds. But in any case, FW regard all forms of probabilism, including bare probabilism, as committed to a position they call `$\rho$-universalism' --- literally, the view that $\rho(t)$ at all times does after all give the correct expectation values for at least macroscopic observables --- which they go on to criticize.

Their criticism is based on a consideration of fluctuations, for which they consider two possible interpretations. The first is that the fluctuations represent probabilistic expectations of how \emph{the same} system's macrovariables will fluctuate over time. As they correctly point out, this is impossible if there are conserved quantities such that the distribution $\rho$ assigns different probabilities to different values of those quantities: in this situation, we would predict that some conserved quantity $C$ has nonzero probability of having value $c$ at time $t$, and also nonzero probability of having value $c'\neq c$ at time $t'$, but (say FW) these are incompatible claims. 

On the second interpretation, the fluctuations represent probabilistic expectations of how identically-prepared systems will differ in the measured values of macrovariables. FW concede that in this case, analytically $\rho$-universalism holds, but object that this is not the empirical situation we find ourselves in: `in laboratory measurements we observe the \emph{same} system at consecutive times rather than drawing different systems out of an urn at random' (p.122; emphasis theirs).

Exploring how probabilism can be saved from these objections, FW formulate two technical conditions (both expressed with respect to a set of macrovariables). A system satisfies \emph{masking} if the fluctuations for each macrovariable are the same for any time-invariant statistical state $\rho$ (so that unrecognized conserved quantities do not affect macrovariable expectation values). A system satisfies \emph{independence} (tacitly, with respect to a given time-invariant statistical state $\rho$) if for sufficiently large times $t$ and for arbitrary initial statistical state $\rho'$ the expectation value of any macrovariable with respect to the time-evolved $\rho'$ and to $\rho$ coincide to any desired degree of accuracy. FW's conclusion is that probabilism is only justified when one or other of these conditions holds, and that it is non-trivial to establish that they do.

\section{Statistical thermodynamics}\label{statistical-thermodynamics}

The core claim of this paper is fairly simple: bare probabilism by itself suffices as a reductive base for thermodynamics, because thermodynamics itself ought to be interpreted as a statistical theory. That is: those quantities in thermodynamics that have a direct microphysical interpretation, like energy, work, or particle number, should be interpreted as expectation values; other quantities, like thermodynamic entropy, should be interpreted as functions of the probability distribution characterizing a thermodynamic system; the statement that a system is in thermodynamic equilibrium should be interpreted as a probabilistic statement about the system, something along the lines of the expected value of any macrovariable being time-invariant.

In more detail, and more carefully: we can distinguish \emph{statistical thermodynamics}, whose parameters are interpreted probabilistically, from \emph{categorical thermodynamics}, whose parameters are interpreted as describing an actual system. My claim is that statistical thermodynamics is derivable from GSM under the bare probabilism interpretation of the latter. Categorical thermodynamics then follows, approximately and with high probability, from statistical thermodynamics for certain sufficiently large systems, basically via the law of large numbers. (And in non-historical contexts, it's not obvious that we benefit from treating categorical thermodynamics as a \emph{theory} rather than just the Law of Large Numbers applied to statistical thermodynamics.)

The details of any such derivation are beyond the scope of this paper (I discuss the technical details of such a derivation in \cite{wallaceirreversibility} but the important point here is that there can be no barrier of principle in deriving statistical thermodynamics from the bare probabilism reading of GSM. If the claims of thermodynamics are understood as categorical then something like AP becomes a substantive extra assumption, in need of justification or explicit posit. But if thermodynamics is itself statistical, there is no conceptual difficulty in supposing (e.g.) that a claim about the \emph{average} work extractable from a system by cyclical processes might be calculable from a probabilistic description of that system. Similarly, if a system is at thermodynamic equilibrium when the \emph{expected} values of its macrovariables are time-invariant, there is no obstacle to identifying thermodynamic equilibrium with statistical macroequilibrium; indeed, that identification is practically analytic.

Why think that statistical thermodynamics is how modern physics interprets thermodynamics? I find the question somewhat difficult to answer: to me, it seems transparent in practically all the modern literature, at least as far back as \cite{tolman}:
\begin{quote}
[I]t is to be emphasized, in accordance with the viewpoint here chosen, that the proposed methods are to be regarded as \emph{really statistical} in character, and that the results which they provide are to be regarded as true \emph{on the average} for the systems in an appropriately chosen ensemble, rather than as necessarily precisely true in any individual case. (63-4, emphasis in original\footnote{I take Tolman's reference to an `ensemble' as being an artifact of a then-prevalent frequentist reading of statistical probability, rather than as essential in the argument.}).
\end{quote}
FW read the physics literature differently, though: they survey thirty textbooks and find most or all committed to AP, and hence (in their reading) to something beyond Bare Probabilism. I am not sure how substantive the difference in our readings is: of course AP is true, indeed in many cases analytic, on the statistical reading of thermodynamics, so in at least some cases the issue is not whether a textbook asserts AP but whether it is committed to the categorical reading of thermodynamics. But to explore this in all cases would be tedious and in any case not decisive (a pretty clear lesson of 20th century philosophy of science is that textbooks are not definitive guides to scientific practice), so I will proceed
differently, by identifying three major research programs in contemporary thermodynamics which clearly presume the statistical interpretation of thermodynamics, and indeed would be unintelligible without it.
\begin{enumerate}
\item The protection of the Second Law from Maxwell's demon via considerations of the thermodynamic cost of erasure (``Landauer's Principle''), as inaugurated by \citeN{bennett1982}. The main focus of this literature has been on very small thermodynamic systems (like the infamous `one-molecule gas'), for which of course there can be no prospect of a version of the Second Law that holds for each individual case: fluctuations can always lead to spontaneous generation of work. The form of the Second Law being defended in this literature is very clearly and explicitly the expected-value form: even with access to a Maxwell demon, one cannot \emph{in expectation} transform work to heat once erasure costs are considered. And this literature equally clearly relies on the identification of thermodynamic entropy with Gibbs entropy, which as Bennett notes (p.936) `is an inherently statistical concept' that cannot even be expressed as the expected value of some microphysically statable quantity.
\item The revolution in classical thermodynamics begun by Jarzynski's \citeyear{jarzynski} celebrated inequality and Crooks' \citeyear{crooksfluctuationtheorem} closely related fluctuation theorem. The Jarznyski equality takes as its starting point a \emph{statistical} interpretation of the Second Law for a system in contact with a heat reservoir: that the \emph{expected} work performed on a system is greater or equal to its change in free energy,
\be
\langle W \rangle \geq \Delta F.
\ee
Jarzynski then strengthens this result to an \emph{equality} (\iec, equation), again statistical in nature:
\be
\langle \e{ - W/k_B T}\rangle = \e{-\Delta F/k_B T}
\ee
from which the classical inequality follows.
\item In quantum thermodynamics, the `one-shot' program begun by Brand\"{a}o~\emph{et al} \citeyear{brandaoetal} again starts with the interpretation of the Second Law as a statistical \emph{average}, and asks what we can say about work extraction if we instead consider \emph{individual} approaches: a variety of inequalities have been established that strengthen the Second Law in the quantum context.
\end{enumerate}

These are not niche programs in modern physics. \cite{bennett1982} has over 2500 citations\footnote{All citation counts from Google Scholar (scholar.google.com), accessed 6/14/2024.}, and the Bennett-Landauer approach to Maxwell's demon is (disapprovingly) acknowledged as the current orthodoxy in Earman and Norton's \citeyear{EarmanNorton1999} historical review of Maxwell's Demon. \cite{jarzynski} has over 5800 citations; \cite{crooksfluctuationtheorem} has over 2900; the Nobel Prize committee\footnote{Advanced Information, NobelPrize.org. Nobel Prize Outreach AB 2024. \\ https://www.nobelprize.org/prizes/physics/2018/advanced-information/ \\ Accessed 6/14/2024.} lists `the first experimental test of Jarzynski's equality' as one of the major applications of optical tweezers, for which Arthur Ashkin shared the 2018 Physics prize. The single-shot program is much newer; still, (Brand\"{a}o \emph{et al} \citeyearNP{brandaoetal}) has over 750 citations.

None of these programs make any sense without a statistical reading of thermodynamics: in each case they quite explicitly make use of it. And in none of the sources I cite above --- or indeed, in any of the other work I'm familiar with in any of these programs --- is there any suggestion that the statistical reading is in need of defending or indeed that it admits any alternative; in each case, it is simply assumed without question. The authors of these sources cannot be regarded as fringe figures; indeed, some of them are acknowledged leaders of their field.\footnote{For instance: Bennett received the Wolf Prize in 2018 and the Breakthrough Prize in 2023; Jarzynski received the Lars Onsager prize in 2019. These are among the highest honors in physics.} It is reasonable to take their views as expressing orthodoxy in at least some large part of the physics community.

To be sure, in \emph{philosophy} of physics the statistical reading of thermodynamics has been severely criticized; it is varyingly claimed to rely on an incoherent notion of objective classical probability, to make objective thermodynamic facts an epistemic and subject-relative matter, to improperly move our focus from how actual individual systems behave, and more. (Variants of these criticisms are made by, \egc, \citeN{alberttimechance}, \citeN{goldsteinboltzmann}, Callender~\citeyear{Callender1999,Callender2002}. FW cite \cite{Callender1999} approvingly.) 

I find these complaints unpersuasive. Objective probability, however philosophically confusing it might be, is manifestly required for the empirical application of statistical mechanics and in any case looks much better once quantum mechanics is taken into account. The epistemic reading of Gibbsian statistical mechanics is in large part optional and in any case thermodynamics is concerned with our capacities, which might reasonably be influenced by our epistemic situation. And there is no difficulty understanding thermodynamics as making claims about individual systems, provided those claims are understood as probabilistic --- something we are in any case used to from quantum physics. (I expand on this defence in (Wallace~\citeyearNP{wallacegibbsnecessity,wallaceirreversibility}); for other defenses, see (\citeNP{maroneygibbs}, \citeNP{myrvoldbook}, \citeNP{robertsonholygrail}).)

But in any case, whether the statistical reading of thermodynamics is correct or even defensible is beside the point. The question that FW ask, and that I am concerned with here, is `can somebody please say what GSM says?' They, and I, are not concerned with the further question: `is what GSM says correct?' I claim that there is overwhelming evidence that in contemporary physics practice, thermodynamics is understood statistically in the first instance, with categorical claims following, insofar as they do follow, only via large-number statistics, and so if we want to understand how GSM is interpreted in contemporary physics, our understanding needs to take this into account. When it is taken into account, there is no difficulty in bare probabilism for the reductive project of deriving thermodynamics from statistical mechanics.

\section{Fluctuations revisited}\label{fluctuations}

Let's consider again FW's question of how to interpret fluctuations in GSM. Suppose that some system is initially at exact statistical equilibrium --- that is, its statistical state $\rho$ is time-invariant. (The interpretation of the claim that the system's statistical state is $\rho$ will depend on our account of statistical-mechanical probabilities --- it might mean that $\rho$ is a quantum state, or a classical distribution understood as the classical limit of a quantum state~\cite{wallacequantumstatmech}, or $\rho$ might be an objectified credence in the sense of \cite{myrvoldbook}, or it might describe the relative frequencies of microstates in a large collection of systems from which our system has been randomly selected. Nothing in this section hinges on this interpretative question.)

If $X$ is a macrovariable for the system, and we measure $X$ at some time $t_1$, the outcome will be somewhat random, with the probability of a given outcome determined by $\rho$, and following FW  we can express the outcome as a fluctuation: that is, as a difference between the actually-obtained value and the mean value. Since the distribution $\rho$ is \emph{ex hypothesi} time-invariant, the probability distribution over these fluctuations is the same whenever the measurement is made, and indeed if we measure the system at a large number of times $t_1,t_2,\ldots t_N$, the probability distribution over results of the $n$th measurement is the same for all $n$.

Does it then follow that we should expect the \emph{statistics} of that large number of measurements to be given by the probability distribution over single measurements? There is no general reason to assume so: it depends on whether the outcomes of measurements at different times are correlated, and there are perfectly realistic cases where we would expect just this.  FW give some examples of this (p.124) but their discussion might give the impression that only exotic systems have this feature. In fact it can readily be found in systems of interest to mainstream statistical mechanics. I will give two examples.

The first is a single harmonic oscillator, previously in thermal contact with a heat bath (so that its statistical state is initially canonical) but now isolated from that bath. The canonical distribution will correctly describe the probability distribution over (say) joint measurements of its position and momentum, at any single time --- but of course any such joint measurement at one time will completely fix the values of any measurement at any other time, so that a sequence of such measurements will not appear as if drawn randomly from an ensemble described by $\rho$ but will trace out a (randomly initialized) harmonic trajectory. (And, since the oscillator is not a chaotic system, this will continue to be true approximately even if the initial measurement has finite precision.)

The second is a macroscopically, but finitely, large ferromagnet, initially heated to well above its ferromagnetic temperature and then allowed to cool. The canonical distribution will predict that the magnetic field at some specified point in the magnet has a non-zero magnitude and a  random direction. But the direction at one time will almost always coincide with the direction at another time: the multi-time correlation between magnetic field values at any point in the magnet is almost perfect.

In each case, the existence of multi-time correlations is not something additional to GSM, but something that can be, and routinely is, calculated within GSM. Given a macrovariable $O$, let $O(t)$ be the macrovariable describing the value of $O$ after the system has evolved under its Hamiltonian dynamics for time $t$. Formally,
\be
O(t) = \exp([\cdot,H]t)O
\ee
where $[,]$ is the Poisson bracket or $-i$ times the commutator, as appropriate.(In QM this can be rewritten as 
\be
O(t)=\e{iHt}O\e{-iHt}
\ee
and we can recognize it as the Heisenberg-picture version of $O$.) Then the two-time correlation function 
\be
\langle O(t)O(0)\rangle_\rho - \langle O(t)\rangle_\rho \langle O \rangle_\rho  
\ee
calculated on the equilibrium distribution $\rho$ is just one more equilibrium expectation value. If it decreases to $\sim 0$ in a time $\tau$, then measurements of $O$ separated by time $\tau$ can be expected to be uncorrelated and their statistical distribution will match the single-time probability distribution of $O$. If it remains large at time $\tau$, those measurements will be correlated and their statistics will not match that distribution.

There is another, equivalent, way to look at this. Suppose we do measure $O$ at time 0 and get result $o$. The correct probability distribution to use to describe the system with respect to subsequent measurements is no longer $\rho$ but $\rho$ conditionalized on $O=o$: that is, we should update the distribution to
\be
\rho'(x) = \mc{N} \rho(x) \delta(O(x)-o)
\ee
with $\mc{N}$ a normalization constant. This will in general not be a macroequilibrium distribution; that is, the measurement has moved the system out of statistical equilibrium. The condition for a measurement at a later time $\tau$ to have probabilities given by the equilibrium distribution $\rho$ is that the system has returned to macroequilibrium: that is, it needs to have an equilibration timescale small compared to $\tau$. And we can identify the two-time correlation function's behavior as telling us exactly this equilibration timescale, at least with respect to that macrovariable.

(It is striking that a multi-time correlation function evaluated at equilibrium tells us about the rate at which a non-equilibrium system equilibrates. This observation is one of the foundational principles of non-equilibrium statistical mechanics, encoded in the Onsager regression hypothesis and the fluctuation-dissipation theorem. See, \egc, \cite{kubononequilibriumbook} for further discussion.)

When a system does not equilibrate (or at any rate does not do so quickly enough) the probability distribution over an observable is not measurable by repeated measurements of that same observable on a single system. To measure that distribution, we would instead need to carry out single measurements on a number of identically prepared systems.

FW consider this possibility, but raise two objections. Firstly, they say, ``[t]he systems produced in this way would approximate the Gibbsian ensemble only if the process of equilibrium preparation was such that the systems ended up being produced according to the measure of the ensemble'', and, they claim, ``there is not reason to believe'' that this will be true. Secondly, they describe the approach as ``of questionable legitimacy because in doing so one gives up the aim of describing the evolution of single systems''.

Neither objection is persuasive. In the first place, of course it is true that we can only measure the equilibrium probability distribution $\rho$ through measurements of many systems if those systems have been prepared such that their statistics are given by that distribution. But that is just to say that we can only do so if the systems really are correctly described by $\rho$, and whoever thought otherwise? When systems equilibrate quickly, of course more or less any preparation process will do, but if they don't, we need to pay attention to the way the system is prepared, to make sure it actually has been prepared in statistical equilibrium. In the two cases I described above (the single particle and the ferromagnet) I was careful to describe their preparation to ensure this: the single particle was extracted from an equilibrated collection of many particles; the ferromagnet was cooled from a temperature at which it would have quickly equilibrated. Alternative preparation processes might not produce systems in statistical equilibrium and in this case the equilibrium distribution will not give the right answers. If I prepared a collection of ferromagnets by magnetizing them all and then giving them to a toddler to play with, probably this will not produce a collection of systems at statistical equilibrium (maybe she likes lining things up).

As for FW's second concern: there is no need to give up on describing the evolution of single systems, only on the idea that the statistics of repeated measurements of \emph{non-equilibrating} systems measure the single-time equilibrium distribution. If we want to study systems that don't quickly equilibrate --- including their time evolution --- we need to study lots of copies of those systems, and to make sure that they start off at statistical equilibrium.  

And indeed this is exactly what is done in modern nanoscale thermodynamics. Microscopic systems --- individual strands of DNA, say --- are allowed to equilibrate with a thermal environment; then the experimenter intervenes on them, and the statistics of their responses are measured and compared with the probabilistic predictions. This has become a major area of experimental thermodynamics in the 21st century, driven by theoretical advances like the Jarzynski equality and technological developments like optical tweezers. Put bluntly: if this approach is of questionable legitimacy, someone needs to tell the 2018 Nobel Physics Prize committee.

Nonetheless FW are clearly correct that a \emph{large part} of statistical thermodynamics is concerned with situations where a single system is repeatedly measured; I turn now to this case.

\section{Gibbsian equilibration}\label{equilibration}

In FW's discussion of repeated measurements of the same system, they worry that there are no general reasons to expect those repeated measurements to give statistics matching the synchronic probability distribution. We have seen that this worry is well-founded for some physically-realistic systems, including microscopic and broken-symmetry systems. Still, we can ask (and FW do ask): why expect \emph{any} system to display these statistics on repeated measurement? As they correctly note: any system with conserved quantities other than energy will display multi-time correlations in repeated measurements of those quantities; if any macrovariable's probability distribution conditional on one value of the conserved quantity differs from its distribution conditional on another variable, then inevitably there will be persistent multi-time correlations in measurements of that quantity.

One class of such conserved quantities can be set aside as harmless. If there are \emph{macrovariables} other than energy that are conserved, it is well known that (as I noted in my presentation of GSM in section \ref{formalism}) the appropriate choice of equilibrium distribution must take that into account. Particle number, for instance, is normally conserved; allowing for this fact leads us to use either a generalization of the canonical ensemble which is determined by the expectation values of both energy and particle number, or a generalized microcanonical ensemble that is uniform on a thin shell of states around a given energy and particle number. In general, the space of systems at equilibrium is parameterized by energy \emph{and any other conserved macrovariables}, both in statistical mechanics and in thermodynamics.

But what if there is a conserved quantity that is \emph{not} a macrovariable, and yet which is such that some macrovariables' values are correlated with it? Then, by definition, the system will not equilibrate: its statistical state will not evolve into a macroequilibrium state. In other words, the problem FW identify is just the problem of establishing that systems which we think `ought' to equilibrate actually do. The properties of masking and independence which they discuss are just aspects of the definition of macro-equilibration.

Proving equilibration is notoriously difficult, and for a century the standard approach has been to provide heuristic arguments as to why equilibration should be expected and then to proceed with statistical mechanics as if it were correct. One could dispute that strategy, but that dispute does not seem particularly relevant to the interpretation of GSM. (Nor is a Boltzmannian approach to statistical mechanics any better off here. On the Boltzmannian conception of equilibration, a ``macrostate'' is a region of phase space where the macrovariables are approximately constant, and the ``equilibrium macrostate'' is the largest such macrostate. Microscopic conserved quantities, if there are any, can cause a system to get stuck in the ``wrong'' macrostate just as easily as they can cause a system to display the ``wrong'' statistics.)

That said, the last 30 years have seen very substantial progress in our understanding of equilibration, mostly in the quantum rather than classical regime. A full review would be far beyond the scope of this paper (see, \egc, (d'Alessio \emph{et al}~\citeyearNP{dalessioreview}) and references therein), but to give one important example: the ``eigenstate thermalization hypothesis'' (\citeNP{deutschETH,srednicki-thermalisation}) relates equilibration to certain statistical features of the distribution of a large system's energy eigenstates --- statistical features which provably characterize `typical' Hamiltonians (for an appropriate definition of `typical') and which can be checked for specific systems via numerical simulation. Equilibration is a difficult problem, but it is not an intractable one.

\section{Conclusion}\label{conclusion}

Gibbsian statistical mechanics assigns a probability distribution $\rho(t)$ to a system at time $t$, and its interpretation is nothing more or less than that this probability distribution is, indeed, the statistical state of the system: that is, it determines the probability distribution over, and so the expectation value of, any variable for the system. But a great deal follows from this. GSM determines the multi-time correlation functions that describe both fluctuations at equilibrium and the approach to equilibrium. It determines the control theory of interventions on a statistical-mechanical system that is statistical thermodynamics: a theory which establishes limits on the expectation values of he work extractable from a system through various control operations. It provides the foundation for the remarkable recent work in thermodynamics that goes beyond these limits and actually provides sharp predictions of expectation values in thermodynamic change. And, through the law of large numbers, it provides a microphysical grounding for the categorical thermodynamics of large systems.

That's what Gibbsian statistical mechanics says. 

\section*{Acknowledgements}

Thanks to Roman Frigg, Charlotte Werndl, and Wayne Myrvold for helpful discussions.


\end{document}